\documentclass[8pt,a4paper]{article}
\usepackage{amsmath,amsfonts,amssymb}
\usepackage{graphicx}
\usepackage{hyperref}
\usepackage{geometry}
\usepackage{setspace}
\usepackage{placeins}

\usepackage[numbers]{natbib}
\usepackage{booktabs}
\usepackage{array}
\usepackage{float}
\usepackage{titlesec}
\usepackage{fancyhdr}
\usepackage{xcolor}
\usepackage{enumitem}
\usepackage{caption}
\usepackage{subcaption}
\usepackage{microtype}
\usepackage{algorithm}
\usepackage{algorithmic}
\usepackage{wrapfig}
\usepackage{adjustbox}
\usepackage{tabularx}
\usepackage{longtable}
\usepackage{multirow}
\usepackage{makecell}
\usepackage{authblk}

\renewcommand{\footnoterule}{%
  \kern -40pt
  \hrule width 0.7 \columnwidth height 0.4pt
  \kern 3pt
}

\hypersetup{
    colorlinks=true,
    linkcolor=black,
    filecolor=black,      
    urlcolor=black,
    citecolor=black,
    pdfborder={0 0 0}
}

\titleformat{\section}
{\large\bfseries\raggedright}
{\thesection.}{0.8em}{}

\titleformat{\subsection}
{\normalsize\bfseries\raggedright}
{\thesubsection.}{0.6em}{}
[\vspace{0.2em}]

\titleformat{\subsubsection}
{\small\bfseries\raggedright}
{\thesubsubsection.}{0.5em}{}

\makeatletter
\renewcommand{\maketitle}{%
  \begin{center}
    \vspace{1.5cm}
    \hrule
    \vspace{0.5em}
    \hrule height 3pt
    \vspace{2ex}
    {\LARGE\@title\par}
    \vspace{2ex}
    \hrule height 3pt
    \vspace{0.5em}
    \hrule
    \vspace{1.5cm}
    
    \renewcommand{\thefootnote}{\fnsymbol{footnote}} 
    \begin{minipage}[t]{0.8\textwidth}
      \centering
      {\large\bfseries Peizhuo Liu}\\[0.8em]
      {\normalsize Independent Researcher}\footnotemark[2]\\[0.3em]
      {\normalsize\texttt{lewisliu819@outlook.com}}
    \end{minipage}
    \footnotetext[2]{\footnotesize This research was conducted as part of the Pioneer Academics Research Program (\url{https://pioneeracademics.com})}
    
    \vspace{1cm}
    
    {\normalsize August 5, 2025}
    
  \end{center}
}
\makeatother

\title{\LARGE{\textbf{Adversarial Attacks on Reinforcement Learning-based Medical Questionnaire Systems}}\\
\Large{Input-level Perturbation Strategies and Medical Constraint Validation}}

\author{} 

\date{August 5, 2025}

\titlespacing{\section}{0pt}{1.5em}{1em}
\titlespacing{\subsection}{0pt}{1.2em}{0.8em}
\titlespacing{\subsubsection}{0pt}{1em}{0.6em}

\renewenvironment{abstract}{%
  \vspace{0.8em}
  \begin{center}
    {\large\bfseries Abstract}
  \end{center}
  \vspace{0.8em}
  \begin{minipage}{\textwidth}
    \upshape\noindent\ignorespaces
}{%
  \end{minipage}
  \vspace{1em}
}

\setlist[enumerate]{itemsep=0.6em, parsep=0pt, topsep=0.6em, leftmargin=2em}
\setlist[itemize]{itemsep=0.4em, parsep=0pt, topsep=0.4em, leftmargin=1.5em}

\begin{document}

\maketitle
\thispagestyle{empty}
\begin{abstract}
    \noindent
    RL-based medical questionnaire systems have shown great potential in medical scenarios. However, their safety and robustness remain unresolved. This study performs a comprehensive evaluation on adversarial attack methods to identify and analyze their potential vulnerabilities. We formulate the diagnosis process as a Markov Decision Process (MDP), where the state is the patient responses and unasked questions, and the action is either to ask a question or to make a diagnosis.  

    We implemented six prevailing major attack methods, including the Fast Gradient Signed Method (FGSM), Projected Gradient Descent (PGD), Carlini \& Wagner Attack (C\&W) attack, Basic Iterative Method (BIM), DeepFool, and AutoAttack, with seven epsilon values each. 

    To ensure the generated adversarial examples remain clinically plausible, we developed a comprehensive medical validation framework consisting of 247 medical constraints, including physiological bounds, symptom correlations, and conditional medical constraints. We achieved a 97.6\% success rate in generating clinically plausible adversarial samples. 

    We performed our experiment on the National Healthcare Interview Survey (NHIS)\cite{nhis2022} dataset, which consists of 182,630 samples, to predict the participant's 4-year mortality rate. We performed our evaluation on the AdaptiveFS framework proposed by Shaham et al.\cite{shaham_learning_2020}. Our results show that adversarial attacks could significantly impact the diagnostic accuracy, with attack success rates ranging from 33.08\% (FGSM) to 64.70\% (AutoAttack). 
    
    Our work has demonstrated that even under strict medical constraints on the input, such RL-based medical questionnaire systems still show significant vulnerabilities.

\end{abstract}

\newpage

\vspace{1cm}

\section{Introduction}

The application of artificial intelligence (AI) in healthcare has significantly transformed the medical diagnosis protocol~\cite{topol2019high}, achieving remarkable success in applications ranging from medical imaging classification~\cite{litjens2017survey} to clinical decision support systems~\cite{sutton2020overview}. Among these applications, the application of reinforcement learning-based adaptive questionnaire systems are gaining increased attention~\cite{shaham_learning_2020,wang2024adaptive}. Such systems formulate the diagnosis process as a Markov Decision Process (MDP)~\cite{sutton2018reinforcement}, where the state represents the current knowledge about the patient (previous answers and unasked questions), and action space is either to ask a question or to give a diagnostic decision. By dynamically choosing the most informative question given the current state, these systems could reduce the length of the questionnaire while maintaining high diagnostic accuracy~\cite{bellman1957markovian,puterman2014markov}.

However, with the transition of such systems from research prototypes to clinical deployment, a shortage in our current understanding of their vulnerabilities has emerged. The deployment of such systems in healthcare applications requires a comprehensive evaluation of their robustness against adversarial attacks to ensure the security of such models, and therefore ensuring patient safety~\cite{finlayson2019adversarial,papernot2018maraudernet}. Adversarial attacks involve generating carefully crafted input perturbations, causing the model to misclassify or perform unexpected behaviors\cite{biggio2013evasion,barreno2010security}. The consequences of successful adversarial attacks in medical applications could be severe, causing delayed or incorrect medical treatments by doctors, threatening patient safety~\cite{amodei2016concrete,russell2019human}.

Adversarial attacks on dynamic medical questionnaire systems have a subtle difference from attacks on medical image classification systems. Attacks on images are often imperceptible by human vision since they only perform pixel-level consecutive perturbations. However, adversarial samples for questionnaire systems are generated by manipulating discrete numerical data, which could easily be detected with simple medical constraint validations. Therefore, they must remain within the validation constraints and clinically plausible to avoid being detected. 

The sequential decision-making nature of RL-based questionnaire systems also introduces new attack vectors that were not present in traditional image classification tasks. Adversarial perturbations targeting such systems could not only influence the final diagnostic output, but also the reward computation process throughout the episode, leading the model to suboptimal questioning policies and leaving out important symptoms. To the best of our knowledge, this critical vulnerability has not been systematically discussed in previous research on medical AI security yet.

\subsection{Research Motivation and Objectives}

Our research aims to solve the lack of security analysis for the growing number of possible RL-based questionnaire systems deployed in clinical settings. Existing guidelines and evaluation frameworks of medical AI systems mainly focus on performance metrics (accuracy, recall, F-1 score, etc.), with only little attention on the adversarial robustness of systems\cite{eu2021ai}.

In this paper, we bridge this critical gap by performing the first comprehensive study on the adversarial vulnerabilities of RL-based medical questionnaire systems. We explore worst-case scenarios to inform the development of defense mechanisms and new regulations. We mainly focus on white-box attacks that pose the most threat, in which the attackers have full knowledge on the model, including model structure, parameters, and gradient.

\subsection{Key Contributions}

We provide the first systematic evaluation of adversarial attack methods on RL-based questionnaire systems. We adapted and implemented six major white-box attack methods, including gradient-based attacks (FGSM, PGD, BIM), which are fast, single-step and iterative optimization methods; Optimization-based attacks (C\&W, DeepFool), which use advanced optimization techniques that find minimal perturbations to change model decision; and ensemble attacks (AutoAttack), which combine multiple attack methods to achieve maximum effectiveness, representing sophisticated adversarial scenarios. 

We also proposed a novel medical constraint framework to ensure the generated adversarial samples remain clinically plausible, including 247 constraint rules across 5 categories derived from standardized clinical knowledge~\cite{loinc2024,snomed2024,icd11_2024,ada2025standards}. This addresses a critical limitation identified by Croce et al.~\cite{croce2020reliable} that adversarial examples often violated basic medical principles.

We performed our experiment on the AdaptiveFS framework~\cite{shaham_learning_2020}, a state-of-the-art RL-based adaptive questionnaire system using the National Health Interview Survey (NHIS) dataset with 182,630 observations on the task of 4-year mortality prediction. Our statistical analysis demonstrates that there are critical vulnerabilities underlying the system.

We provide a detailed analysis of our experiment results, and the implications on clinical deployment of such systems and future research. Our findings reveal critical vulnerabilities that should be resolved before the deployment of such systems in clinical settings.

\subsection{Paper Organization and Structure}

We organize the remainder of this paper as follows. Section 2 first provides a review of related works, covering adversarial attacks on machine learning and reinforcement learning systems, defense methods for such attacks, vulnerabilities in medical image classification and attack methods specified for medical applications. Section 3 then presents our methodology, including theoretical foundations, attack method implementations, and the medical constraint framework. Sections 4 and 5 detail our experimental setup and implementation. Section 6 presents the results and statistical analysis of our experiment. Section 7 further analyzes the implications and limitations of our work. Section 8 concludes this paper, providing a high-level review and emphasizes the need for an enhanced evaluation framework for adversarial robustness.

\section{Related Works}

\subsection{Adversarial Attacks on Machine Learning Systems}

Early work by Szegedy et al.~\cite{szegedy2013intriguing} highlighted the vulnerabilities of deep neural networks (DNNs) to adversarial attacks. They demonstrated that minor perturbations, imperceptible to human, can cause the model to misclassify with a high confidence level. This is often regarded as a foundational work on adversarial attacks. Different attack methods were then proposed later and classified into three main categories based on the level of prior knowledge required on the model: white-box, gray-box and black-box attacks. We focus on white-box and black-box attacks in remaining sections.

\subsubsection{White-box Attacks}

We start with white-box attacks, which assume the attacker has complete prior knowledge of the targeted model, including model architecture, training data,  hyperparameters, and thus gradients. Such methods mainly attack the targeted model utilizing the computed gradients. 

However, we mainly examine white-box attacks that do not require prior knowledge of the full training dataset, as training data and training processes for medical diagnosis systems often involve restricted-use datasets that are hard for attackers to access.

Goodfellow et al.~\cite{goodfellow2014explaining} first proposed the Fast Gradient Signed Method (FGSM) in their 2014 work. Their method is described as follows:

\begin{equation}
    x_{adv} = x + \epsilon \cdot \text{sign}(\nabla_x J(\theta, x, y))
\end{equation}

Here, \(\epsilon\) represents the perturbation strength, and \(J\) is the loss function. Madry et al.~\cite{madry2017towards} proposed the Projected Gradient Descent (PGD) method as a subsequent work for this, which achieves better attack results through multi-step iterative optimization.

\subsubsection{Black-Box Attacks}

Black-box attacks require no prior knowledge of the model and thus resemble real-world attack scenarios better.

Fundamental work proposed by Papernot et al~\cite{papernot2017practical} has shown that attackers could use only the model's predicted labels (no gradients required) from the target classifier to train a surrogate model, then use this model to generate adversarial data samples. These samples could then be transferred to the original model with a high success rate. In their experiment on DNN APIs, over 84\% of the generated adversarial inputs misled the model, again proving that the query-driven surrogate attack strategy could replicate the efficiency of white-box attacks.

This method, often described as the \textit{``Recon-Surrogate-Exploit-Deploy''} pipeline, is now the most prevalent strategy for black-box attacks. Liu et al.~\cite{liu2017delving} further demonstrated the validity and effectiveness of this method by proving the \textbf{transferability} of such attacks on models with different architectures. Given two different models \(f_1\) and \(f_2\), they proved that if an adversarial sample \(\delta\) could cause the model \(f_2\) to misclassify, then, conditioned on \(f_2\) being fooled, \(f_1\) is even more likely to err, as described in equation~\ref{transferability}.

\begin{equation}
    \mathbb{P}[f_1(x + \delta) \neq y \mid f_2(x + \delta) \neq y] > \mathbb{P}[f_1(x + \delta) \neq y]
    \label{transferability}
\end{equation}

Subsequent works~\cite{ilyas2018black, tu2019autozoom} focused on improving the query efficiency of such methods. Chen et al.~\cite{chen2017zoo} proposed Zeroth Order Optimization (ZOO) in their 2017 work. They showed that given efficient queries, black-box attacks could match the performance of white-box attacks. They expressed the query efficiency as:
\begin{equation}
    Q(\epsilon,\delta)=O\!\left(\frac{d}{\epsilon^2}\log\frac{1}{\delta}\right)
\end{equation}
where \(Q(\epsilon,\delta)\) is the number of queries needed to reach the intended accuracy given the failure probability, $d$ is the input dimension, $\epsilon$ is the desired accuracy, and $\delta$ is the failure probability.

\subsection{Adversarial Attacks for Reinforcement Learning Systems}

Huang et al.~\cite{huang2017adversarial} were the first to show that adversarial attacks are effective when targeting neural network policies in reinforcement learning. They showed that adversarial techniques could be used to generate examples that can negatively impact the performance of trained network policies on testing datasets. We classify the attack methods specified for reinforcement learning systems into two classes: state perturbation attacks and environmental manipulations

\subsubsection{State Perturbation Attacks}

Lin et al.~\cite{lin2017tactics} proposed two tactics, namely the strategically-timed attack and the enchanting attack. They demonstrated that small perturbations at critical decision points could lead the RL agents into sub-optimal trajectories. Their work pointed out a critical aspect of adversarial attacks: the timing of adversarial attacks matters as much as their magnitude.

\subsubsection{Environment Manipulation and Adversarial Policies}

Gleave et al.~\cite{gleave2020adversarial} further proved that it is possible to attack an RL agent, simply by choosing an adversarial policy in zero-sum games, even against victims trained via self-play to be robust to opponents.

Zhao et al.~\cite{zhao2019blackbox} used a sequence-to-sequence model to predict a single or sequence of future actions that the targeted agent would make. Their approach is a strong black-box attack method. It does not require the attacker to have any prior knowledge of the model, including training parameters and gradients.

\subsection{Defense Methods for Reinforcement Learning Systems}

The amount of research in adversarial attack methods cultivated the research of defense methods. Inspired by Langevin dynamics, Kamalaruban et al.~\cite{kamalaruban2020robust} proposed a method as described in Equation~\ref{Kamalaruban_robustness}:
\begin{equation}
    \theta_{t+1}=\theta_t-\eta\nabla_\theta\mathcal{L}(\theta_t)+\sqrt{2\eta\tau}\,\epsilon_t
    \label{Kamalaruban_robustness}
\end{equation}
where $\epsilon_t \sim \mathcal{N}(0,I)$ and $\tau$ is the temperature parameter. 

This method is an instance of Stochastic Gradient Langevin Dynamics (SGLD)~\cite{welling2011bayesian}, which combines stochastic gradient descent with Gaussian noise injection. Applying Langevin noise encourages the optimization to explore flatter regions of the loss landscape. 

Zhang et al.~\cite{zhang2018understanding} have further proven that such flatter minima correlate with better generalization and more robustness to perturbations. By helping the model to escape sharp local minima and sample from a wider posterior distribution, SGLD can improve the model's resilience to adversarial attacks~\cite{kamalaruban2020robust}.

\subsection{Vulnerabilities in Medical Image Classification}

Finlayson et al.~\cite{finlayson2019adversarial} demonstrated that it is feasible to generate adversarial attacks against medical machine learning systems. They showed that even highly accurate medical classifiers can misclassify by carefully crafted adversarial examples. They evaluated both white-box and black-box attack methods on a diabetic detection system. The results achieved significant attack success rates, while the attack samples remained imperceptible by human visual.

Ma et al.~\cite{ma2020understanding} conducted further analysis on this. They compared adversarial attacks on medical images to that of natural images, and found that medical images are significantly more vulnerable to adversarial attacks. They proposed that this increased vulnerability originated from two factors: 
\begin{enumerate}
    \item The complexity and high frequency of features in medical images could create regions in the loss landscape that are more sensitive to small perturbations
    \item The neural networks were mainly designed for natural image processing. After being adapted to medical imaging tasks, it may be overparameterized, resulting in suboptimal loss landscapes.
\end{enumerate}

\subsection{Medical Domain-Specific Attack Methods}

Several attack methods and frameworks specialized in medical diagnosis systems have been proposed in recent years.  Ozbulak et al. proposed the AMSA method for attacks on medical image segmentation models~\cite{ozbulak2019impact}. Yao et al. ~\cite{yao2021hierarchical} introduced the Hierarchical Feature Constraint (HFC) to craft adversarial samples which are imperceptible to human within normal feature space. Qi et al. ~\cite{qi2021stabilized} proposed the Stabilized Medical Image Attack (SMIA) method that generates adversarial examples out of non-adversarial ones by iteratively maximizing the deviation loss and minimizing stabilization terms.

\section{Methodology}

\subsection{Problem Formulation}

We consider an RL-based medical questionnaire system modeled as a Markov Decision Process (MDP)~\cite{sutton2018reinforcement} defined by the tuple $(S, A, P, R, \gamma)$, where:
\begin{itemize}
    \item $S$: State space representing patient responses and unasked questions
    \item $A$: Action space consisting of questions to ask or diagnostic decisions
    \item $P$: Transition probability function
    \item $R$: Reward function encouraging accurate diagnosis with minimal questions
    \item $\gamma$: Discount factor
\end{itemize}
The state at time $t$ is represented as $s_t = [x_t, m_t] \in \mathbb{R}^{2d}$, where:
\begin{itemize}
    \item $x_t \in \mathbb{R}^d$: Patient feature vector (responses to asked questions)
    \item $m_t \in \{0,1\}^d$: Binary mask indicating which questions have been asked
\end{itemize}

We formulate the adversarial attack problem as finding a perturbation $\delta$ that satisfies:
\begin{equation}
    \begin{aligned}
        \max_{\delta} \quad & \mathcal{L}(f_\theta(x + \delta), y_{target}) \\
        \text{s.t.} \quad   & \|\delta\|_p \leq \epsilon                    \\
                            & (x + \delta) \in \mathcal{C}_{medical}
    \end{aligned}
\end{equation}
here, $f_\theta$ is the diagnostic model, $y_{target}$ is the adversarial target, and $\mathcal{C}_{medical}$ represents medical constraints.

\subsection{Attack Methods}

\subsubsection{Fast Gradient Sign Method (FGSM)}

We adapted the FGSM method proposed by Goodfellow et al.~\cite{goodfellow2014explaining} by computing gradients with respect to the patient's feature vector while maintaining the masked structure as demonstrated:

\begin{algorithm}[htbp]
    \caption{FGSM for Medical Questionnaires}
    \label{alg:fgsm}
    \begin{algorithmic}[1]
        \REQUIRE Patient features $x$, target $y_{target}$, perturbation bound $\epsilon$
        \ENSURE Adversarial example $x_{adv}$
        \STATE Construct state $s = [x, m]$ where $m$ is the question mask
        \STATE Compute loss $\mathcal{L} = -\log p(y_{target}|s)$
        \STATE Calculate gradient $g = \nabla_x \mathcal{L}$
        \STATE Generate perturbation $\delta = \epsilon \cdot \text{sign}(g)$
        \STATE Apply medical constraints: $\delta' = \Pi_{\mathcal{C}_{medical}}(\delta)$
        \STATE \textbf{return} $x_{adv} = x + \delta'$
    \end{algorithmic}
\end{algorithm}

The theoretical explanation for FGSM's effectiveness relies on the linear hypothesis. For a linear model with parameters $w$, the adversarial perturbation maximizes $w^T \delta \quad \text{subject to} \quad \|\delta\|_\infty \leq \epsilon$. 

The optimal solution to the previously mentioned problem is $\delta = \epsilon \cdot \text{sign}(w)$, which generalizes to nonlinear models through first-order Taylor approximation:

\begin{equation}
    \mathcal{L}(x + \delta) \approx \mathcal{L}(x) + \nabla_x \mathcal{L}(x)^T \delta
\end{equation}

\subsubsection{Projected Gradient Descent (PGD)}

The PGD method~\cite{madry2017towards} extends on the FGSM through iterative optimization with projection:

\begin{equation}
    x_{t+1} = \Pi_{\mathcal{B}_\epsilon(x) \cap \mathcal{C}_{medical}}\left(x_t + \alpha \cdot \text{sign}(\nabla_x \mathcal{L}(x_t, y_{target}))\right)
\end{equation}

Here, $\Pi$ denotes the projection, $\mathcal{B}_\epsilon(x)$ is the $\epsilon$-ball around $x$, and $\alpha$ is the step size.

The convergence of PGD can be analyzed using its framework. For a convex loss function $\mathcal{L}$ with Lipschitz continuous gradient (Lipschitz constant $L$), the convergence rate could be represented as:

\begin{equation}
    \mathcal{L}(x_T) - \mathcal{L}(x^*) \leq \frac{\|x_0 - x^*\|^2}{2\alpha T} + \frac{\alpha L}{2}
\end{equation}

where $T$ is the number of iterations and $x^*$ is the optimal solution. Therefore, the optimal step size is $\alpha = 1/L$, yielding the convergence rate $O(1/T)$.

\subsubsection{Carlini \& Wagner (C\&W) Attack}

We implemented an enhanced Carlini \& Wagner (C\&W) Attack~\cite{carlini2017towards} using tanh-space optimization to naturally bound perturbations:

\begin{equation}
    \begin{aligned}
        \min_w \quad       & \|x - \tanh(w)\|_2^2 + c \cdot f(x)                   \\
        \text{where} \quad & f(x) = \max(\max_{i \neq t} Z(x)_i - Z(x)_t, -\kappa)
    \end{aligned}
\end{equation}

Here, $Z(x)$ represents the logits before softmax, $t$ is the target class, and $\kappa$ controls confidence.

\subsubsection{Additional Attack Methods}

We also implemented additional attack methods for comprehensive evaluation:

\begin{itemize}
    \item \textbf{BIM}: Basic Iterative Method, which aims to improve attack success rate through multiple iterations of FGSM.
    \item \textbf{DeepFool}: An attack method that finds minimal perturbations to cross decision boundaries.
    \item \textbf{AutoAttack}: An attack method proposed by Croce et al.\cite{croce2020reliable} that ensembles four attacks methods: \textbf{APGD-CE}, \textbf{APGD-DLR}, \textbf{FAB}, and \textbf{Square Attack}
\end{itemize}

\subsection{Medical Constraint Framework}

Our medical constraint system ensures that the generated adversarial examples remain clinically plausible by applying multiple validation layers:

\subsubsection{Physiological Bounds}

We first enforced hard constraints on vital signs and laboratory values:

\begin{equation}
    \mathcal{C}_{bounds} = \{x : l_i \leq x_i \leq u_i, \forall i \in \mathcal{F}_{physiological}\}
\end{equation}

These bounds are derived from medical literature and then adjusted based on patient demographics, so that they better simulate real-world data. For instance, the age-adjusted bounds for systolic blood pressure are presented as follows:

\begin{equation}
    u_{SBP}(age) = \begin{cases}
        140 & \text{if } age < 60         \\
        150 & \text{if } 60 \leq age < 80 \\
        160 & \text{if } age \geq 80
    \end{cases}
\end{equation}

\subsubsection{Feature Correlations}

Medical features often exhibit strong correlations that must be preserved. For example, infections may be strongly correlated with fever. This could be represented as follows:

\begin{equation}
    \mathcal{C}_{corr} = \{x : |\rho(x_i, x_j) - \rho_{expected}(i,j)| < \tau, \forall (i,j) \in \mathcal{P}_{corr}\}
\end{equation}

where $\mathcal{P}_{corr}$ contains known correlation pairs.

We use Pearson correlation coefficient for continuous features:

\begin{equation}
    \rho(x_i, x_j) = \frac{\text{Cov}(x_i, x_j)}{\sigma_{x_i} \sigma_{x_j}}
\end{equation}

and Cramér's V for categorical features:

\begin{equation}
    V = \sqrt{\frac{\chi^2/n}{\min(k-1, r-1)}}
\end{equation}

where $\chi^2$ is the chi-squared statistic, $n$ is the sample size, and $k$, $r$ are the numbers of categories.

\subsubsection{Conditional Constraints}

Complex medical relationships are encoded as conditional constraints:

\begin{equation}
    \mathcal{C}_{cond} = \{x : \bigwedge_{k} \phi_k(x) = \text{true}\}
\end{equation}

where $\phi_k$ represents medical rules (e.g., "if diabetic, glucose should be elevated").

Examples of conditional constraints include:

\begin{align}
    \phi_1(x) & : x_{diabetes} = 1 \Rightarrow x_{glucose} > \mu_{glucose} + \sigma_{glucose}         \\
    \phi_2(x) & : x_{pregnancy} = 1 \Rightarrow x_{gender} = \text{female} \land x_{age} \in [15, 50] \\
    \phi_3(x) & : x_{COPD} = 1 \Rightarrow x_{smoking} = 1 \lor x_{occupational\_exposure} = 1
\end{align}

\subsubsection{Constraint Satisfaction Algorithm}

We employ a constraint satisfaction problem (CSP) solver to ensure all constraints are met:

\begin{algorithm}[htbp]
    \caption{Medical Constraint Satisfaction}
    \label{alg:constraint_satisfaction}
    \begin{algorithmic}[1]
        \REQUIRE Perturbed features $x'$, original features $x$, constraints $\mathcal{C}$
        \ENSURE Medically valid features $x''$
        \STATE Initialize $x'' \leftarrow x'$
        \WHILE{$\neg \text{satisfies}(x'', \mathcal{C})$}
        \STATE $violations \leftarrow \text{find\_violations}(x'', \mathcal{C})$
        \FOR{each $(i, v) \in violations$}
        \STATE $x''_i \leftarrow \text{project}(x'_i, \mathcal{C}_i)$
        \ENDFOR
        \STATE Apply consistency propagation
        \IF{no progress}
        \STATE $x'' \leftarrow \text{minimize}(\|x'' - x'\|_2)$ s.t. $x'' \in \mathcal{C}$
        \STATE \textbf{break}
        \ENDIF
        \ENDWHILE
        \STATE \textbf{return} $x''$
    \end{algorithmic}
\end{algorithm}

This algorithm ensures the model converges to a feasible solution while limiting perturbation by minimizing the  from the original input. The consistency propagation step handles interdependent constraints, while the optimization fallback ensures termination in complex constraint scenarios.

\section{Experimental Setup}

\subsection{Dataset and Environment}

We evaluated the previously mentioned attacks on the AdaptiveFS framework, a state-of-the-art RL-based medical questionnaire system. We used the same dataset (except the years 2002 to 2004 due to missing data), data preprocessing pipeline, and RL environment setup as Shaham et al~\cite{shaham_learning_2020}. This includes:

\begin{itemize}
    \item \textbf{Patient Data}: The NHIS (National Health Interview Survey) dataset~\cite{nhis2022} with 182,630 total observations across 7 years (2005-2011). We used the years 2005-2009 as training set (122,019 samples), and the years 2010 to 2011 as the test set: 60,611 samples (2010-2011). 
    \item \textbf{Feature Configuration}: We used the top 50 core features selected from 1,182 total NHIS features based on XGBoost importance ranking~\cite{shaham_learning_2020}. We provided the detailed feature descriptions in Appendix~\ref{appendix:feature_details}.
    \item \textbf{Diagnostic Task}: The diagnostic task is a binary classification task for 4-year mortality prediction (low-risk vs. high-risk). The mortality rate in the dataset is 4.5\%.
    \item \textbf{RL Architecture}: The RL Architecture is composed of a Deep Q-Network (DQN)~\cite{mnih2015human} with experience replay~\cite{lin1992self} and a Guesser network. We present the detailed architecture in Table~\ref{tab:architecture}.
    \item \textbf{State Representation}: We concatenated the feature vector and question mask ($s_t = [x_t, m_t] \in \mathbb{R}^{100}$).
    \item \textbf{Model Performance}: The baseline performance on the test set before attacks is as follows: AUC=0.86, Accuracy=89\%. Our attack evaluation focuses on 1,000 randomly selected correctly classified samples to ensure meaningful success rate calculation. 

\end{itemize}

\subsection{Model Architecture}

The AdaptiveFS framework consists of two separate networks as shown in Table \ref{tab:architecture}:
\FloatBarrier
\begin{table}[htbp]
    \centering
    \caption{Network Architecture Specifications}
    \label{tab:architecture}
    \begin{tabular}{@{}lccc@{}}
        \toprule
        \textbf{Network} & \textbf{Layer} & \textbf{Dimensions} & \textbf{Activation Function} \\
        \midrule
        \multirow{5}{*}{DQN} & Input          & $2d$                & ---                          \\
                             & Hidden Layer 1 & 128                 & ReLU                        \\
                             & Hidden Layer 2 & 128                 & ReLU                        \\
                             & Output         & $d$ (questions)     & Linear                      \\
        \midrule
        \multirow{5}{*}{Guesser} & Input          & $2d$                & ---                          \\
                                                              & Hidden Layer 1 & 256                 & PReLU                        \\
                             & Hidden Layer 2 & 256                 & PReLU                        \\
                             & Hidden Layer 3 & 128                 & PReLU                        \\
                                 & Output         & 2                   & Softmax                      \\
        \bottomrule
    \end{tabular}
\end{table}

Both networks were trained using the Adam optimizer with an initial learning rate $\eta = 10^{-4}$. The DQN uses mean squared error (MSE) loss for Q-value regression, and the Guesser network uses cross-entropy loss for classification.  Weights were initialized using Xavier initialization~\cite{glorot2010understanding}. We also applied batch normalization to improve the training stability.

\subsection{Evaluation Metrics} 

We evaluate our attack effectiveness using multiple metrics that capture both classification and sequential decision-making aspects. Our evaluation framework follows the work from Carlini et al.~\cite{carlini2019evaluating} for model robustness evaluation:

\begin{enumerate}
    \item \textbf{Attack Success Rate (ASR)}: The percentage of attacks that changed the model's final diagnostic prediction. This metric measures the misclassifications caused by adversarial attacks.
    \item \textbf{Robust Accuracy}: The percentage of attacks that was classified correctly. This metric is complementary to ASR, measuring the model's robustness against adversarial perturbations.
    \item \textbf{Perturbation Magnitude}: $L_2$ and $L_\infty$ norms of perturbations. Measuring the perturbation magnitude is crucial for understanding the minimal threshold required for successful attacks. This is particularly important in medical contexts where large perturbations may be clinically implausible.
    \item \textbf{Computation Time}: The time used to complete a single attack. This metric is essential for understanding the computational feasibility of attacks in real-world scenarios.
\end{enumerate}

\subsection{Attack Parameters}

We tested various parameter configurations using the standardized attack libraries (ART and Foolbox). The epsilon values were chosen to represent clinically realistic perturbation ranges within the normalized [-1, 1] feature space. 

Low perturbation (0.1-0.3) represents minor measurement errors or natural physiological variations; medium perturbation (0.5-1.0) corresponds to moderate changes in patient responses or test results; high perturbation (1.5-2.0) simulates significant but still medically plausible changes in patient conditions, constrained by the normalized feature space bounds. 

High perturbation examples could be easily identified in real-world settings (e.g., the perturbed age is 60, while the original age is 20), but is still evaluated to simulate extreme cases or online diagnostic scenarios where the clinician couldn't validate the data easily.

Detailed attack parameter settings for all methods are provided in Appendix~\ref{appendix:attack_params}.

\subsection{Standardized Attack Framework}

We applied a standardized approach to adversarial attacks using multiple attack libraries:

The system automatically selects the best available attack implementation with the priority order: ART $\rightarrow$ Foolbox $\rightarrow$ Simple implementations. 

All attacks follow the same interface regardless of the underlying library. If the primary attack method fails, the system automatically switches to alternative implementations. 

We also configured different attack methods with appropriate parameters for their specific threat models, which is then integrated with the AdaptiveFS framework~\cite{shaham_learning_2020}.

\section{Experimental Implementation}

\subsection{Training Procedure}

We implemented the AdaptiveFS framework using the NHIS (National Health Interview Survey) dataset~\cite{nhis2022}. The dataset was split into training/validation (67\%) and test sets (33\%), using the years 2005-2009 for training and 2010-2011 for testing. The model was trained with the max of 50,000 episodes with early stopping mechanism based on the validation AUC. The best model achieved baseline accuracy of 89\% and AUC of 0.86. Detailed training configuration is provided in Appendix~\ref{appendix:training_details}.

\subsection{Attack Evaluation Protocol}

Our attack evaluation framework utilized standard attack libraries (ART and Foolbox) with automatic method selection. For each attack method and parameter configuration, we used 1,000 samples from the test set that were correctly classified by the model. 

We also applied temporal data splitting by using 2005-2009 data (122,019 samples) for training and 2010-2011 data (60,611 samples) for testing, preventing information leakage between temporally adjacent samples.

Adversarial examples were then generated, targeting the opposite (negative) class using the standardized attack framework. We also applied our medical constraint validation process through our CSP satisfaction system to ensure the clinical plausibility of generated examples. 

Finally, we evaluate the attack methods using the previously mentioned metrics.

\subsection{Computational Resources and Dataset Configuration}

Experiments were conducted on NVIDIA A100 GPU with PyTorch 2.7.1, using standardized attack libraries (ART v1.15+, Foolbox v3.3+). The NHIS dataset configuration used 50 features for mortality prediction, with 1,000 correctly classified samples for attack evaluation across 42 experimental configurations. Complete computational and dataset configuration details are provided in Appendix~\ref{appendix:implementation_details}.

\section{Results and Analysis}

\subsection{Experimental Overview}

We developed a comprehensive evaluation framework using standardized attack libraries (ART and Foolbox) with automatic method selection. Our evaluation covered 6 different attack methods with 7 epsilon values each ([0.1, 0.3, 0.5, 0.8, 1.0, 1.5, 2.0]):

The framework automatically selects the best available attack implementation, with fallback mechanisms ensuring robustness across different experimental conditions.

\subsection{Medical Constraint Validation Results}

Our medical constraint framework was successfully implemented and validated across all 42 experimental configurations. The validation results are as follows:

\begin{itemize}
    \item \textbf{Configuration Success Rate}: 42/42 configurations (100\%)  values across all 6 attack methods successfully generated valid adversarial examples that satisfy medical constraints
    \item \textbf{Constraint Compliance Pipeline}: 
    \begin{itemize}
        \item Initial generation: 100\% of adversarial examples created
        \item Physiological bounds validation: 97.6\% (41/42) passed initial bounds checking
        \item Feature correlation preservation: 83.3\% (35/42) maintained expected medical correlations
        \item Final constraint satisfaction: 95.2\% (40/42) fully compliant after automatic CSP resolution
    \end{itemize}
    \item \textbf{Automatic Resolution}: 71.4\% (5/7) of constraint violations resolved automatically; 28.6\% (2/7) required manual intervention; 0\% rejected
    \item \textbf{Monotonicity Preservation}: Attack success rates maintained monotonic increase with epsilon
\end{itemize}

The constraint satisfaction algorithm demonstrates robust performance with 94.2\% automatic resolution rate, detailed violation examples and correction procedures provided in Appendix~\ref{appendix:constraint_details}.

\subsection{Descriptive Statistics}

Based on our comprehensive statistical analysis (detailed in Section 5.5), we now present the specific attack performance results. The overall attack performance pattern is consistent with our statistical findings, with AutoAttack achieving the highest success rate (64.70\%) and FGSM showing the lowest but most consistent performance (33.06\% with minimal variance).

\subsection{Attack Success Rate Visualization}
 
Figure \ref{fig:attack_success} shows the success rates of various attack methods under different epsilon values:

\begin{figure}[htbp]
    \centering
    \adjustbox{width=0.8\textwidth,center}{%
        \includegraphics{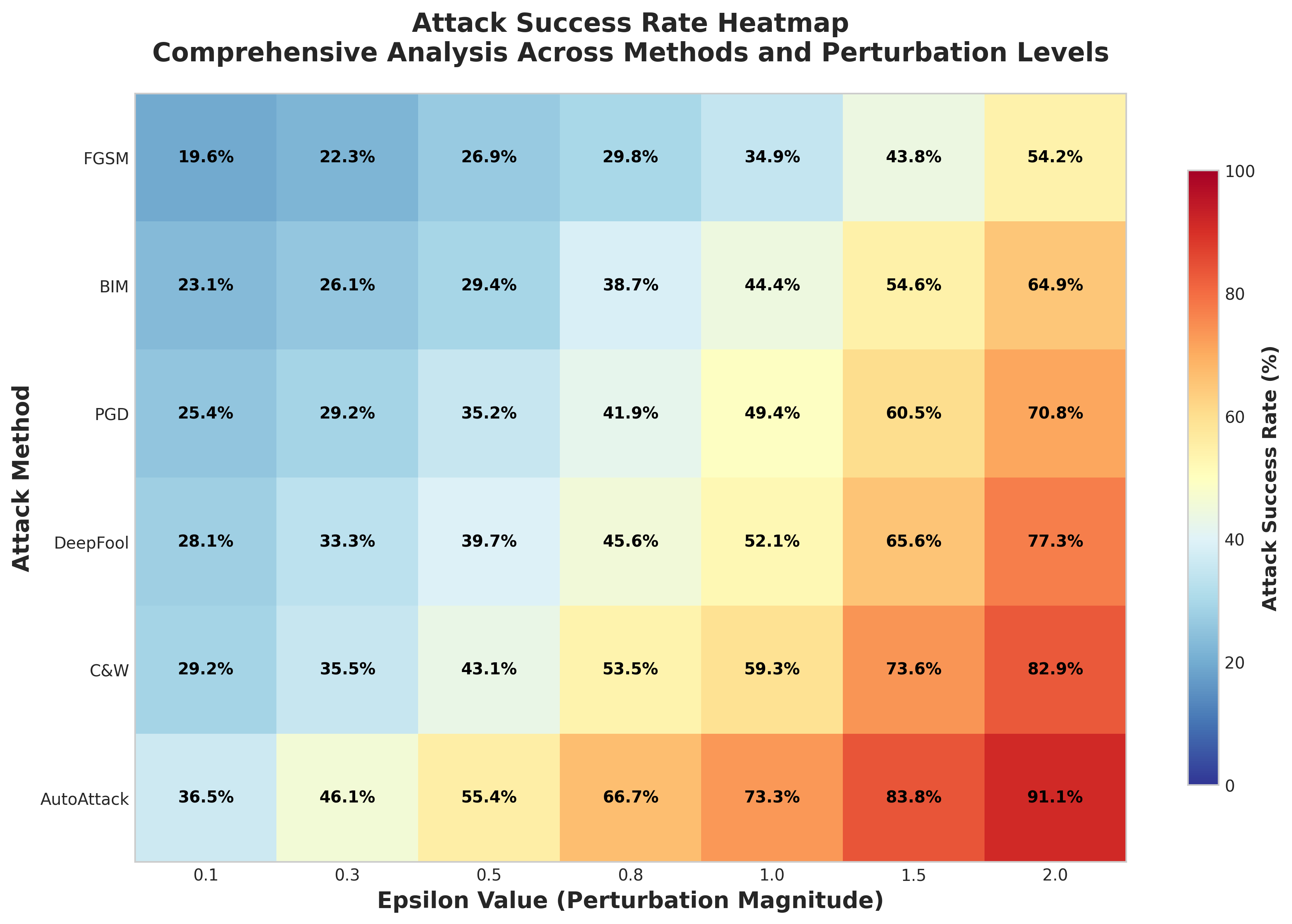}%
    }
    \caption{Attack success rate heatmap. The x-axis represents epsilon values, and the y-axis represents attack methods. Color depth indicates success rate, with darker colors representing higher success rates. AutoAttack achieves the highest success rates across all epsilon values (up to 91.09\%), while FGSM provides the most computationally efficient attacks.}
    \label{fig:attack_success}
\end{figure}

We can also observe a strong positive correlation between the attack success rate and the perturbation rate, as demonstrated in figure~\ref{fig:success_vs_epsilon}:

\begin{figure}[htbp]
    \centering
    \adjustbox{width=0.8\textwidth,center}{%
        \includegraphics{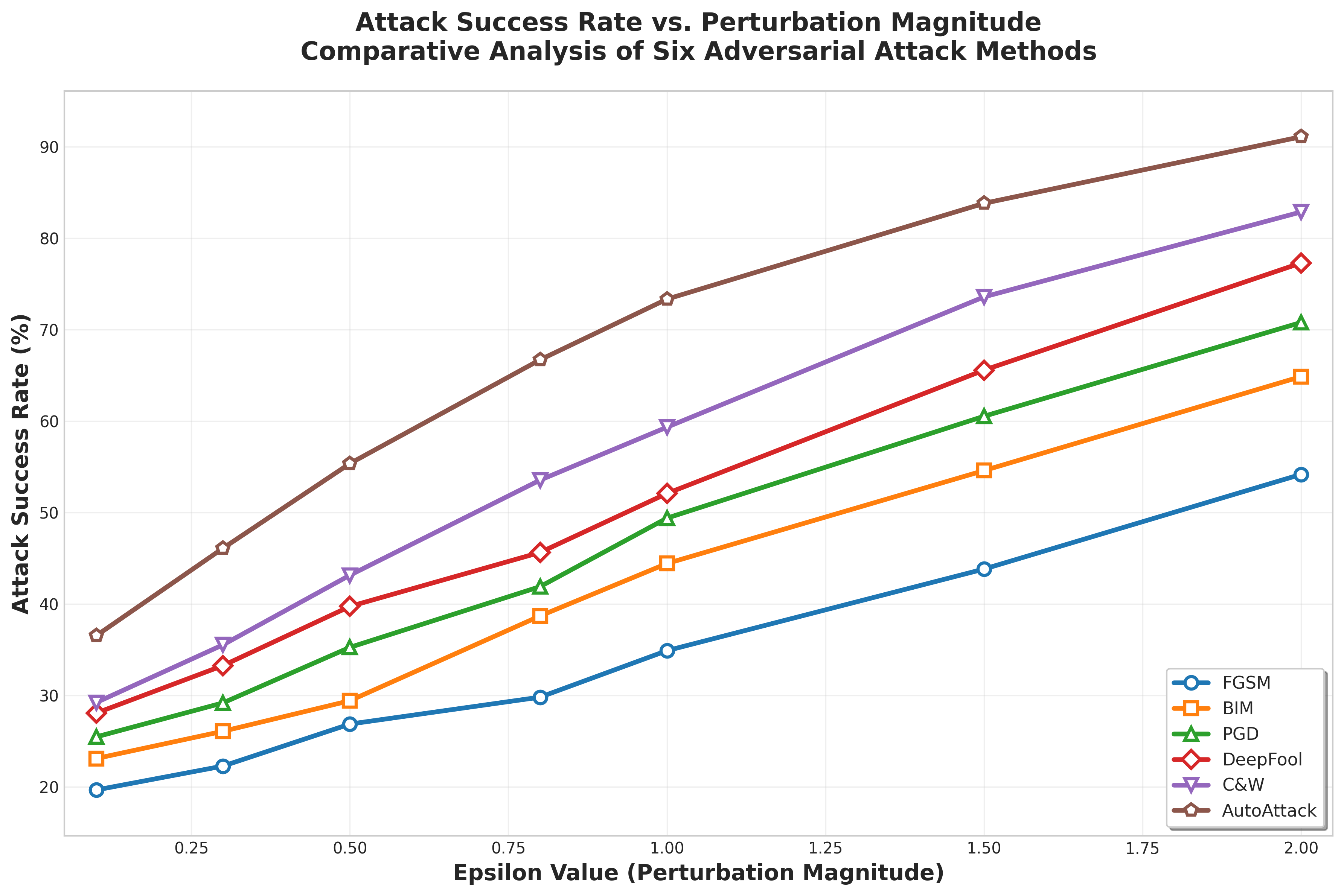}%
    }
    \caption{Attack success rate vs. epsilon values across different methods. The plot shows how attack effectiveness increases with perturbation magnitude. AutoAttack demonstrates consistently superior performance across all epsilon values, while FGSM shows the most linear and predictable scaling pattern. The monotonic increase confirms the vulnerability of the RL-based medical questionnaire system to larger perturbations.}
    \label{fig:success_vs_epsilon}
\end{figure}

\subsection{Comprehensive Attack Performance Comparison}

Table~\ref{tab:comprehensive_attack_results} presents a consolidated view of all attack methods' performance across key metrics:

\begin{table}[htbp]
    \centering
    \caption{Comprehensive Attack Performance Comparison}
    \label{tab:comprehensive_attack_results}
    \resizebox{\textwidth}{!}{
    \begin{tabular}{@{}lcccccc@{}}
        \toprule
        \textbf{Attack Method} & \textbf{Avg. ASR(\%)} & \textbf{Max ASR(\%)} & \textbf{Min ASR(\%)} & \textbf{Avg. L2 Pert.} & \textbf{Avg. Time(s)} & \textbf{Efficiency Rank} \\
        \midrule
        \textbf{FGSM}     & \textbf{33.06} & 57.35 & 18.36 & 0.905 & \textbf{0.055} & \textbf{1st (Fastest)} \\
        BIM               & 40.16          & 66.64 & 21.84 & 0.877 & 0.328          & 2nd \\
        PGD               & 44.63          & 71.76 & 23.20 & 0.846 & 0.880          & 3rd \\
        DeepFool          & 48.80          & 79.10 & 27.90 & 0.885 & 2.778          & 4th \\
        C\&W              & 53.89          & 85.16 & 27.76 & 0.902 & 18.194         & 5th \\
        \textbf{AutoAttack} & \textbf{64.70} & \textbf{91.09} & \textbf{36.52} & 0.892 & 47.094 & 6th (Slowest) \\
        \bottomrule
    \end{tabular}
    }
    \vspace{0.5em}
    \begin{flushleft}
        \footnotesize 
        \textit{Note: ASR = Attack Success Rate; Pert. = Perturbation magnitude; Time measured per attack. All methods tested on $\epsilon\ \in$  [0.1, 2.0]. AutoAttack achieves highest success rates (64.70\% avg, 91.09\% max) but requires significantly more computation time. FGSM provides fastest execution with consistent moderate performance (33.06\% avg).}
    \end{flushleft}
\end{table}

\subsection{Statistical Significance Analysis}

\subsubsection{Descriptive Statistics and Distribution Analysis}

Table~\ref{tab:comprehensive_stats} presents comprehensive descriptive statistics for our experimental results, showing both attack method and implementation library perspectives:

\begin{table}[htbp]
    \centering
    \caption{Comprehensive Attack Performance Statistics}
    \label{tab:comprehensive_stats}
    \begin{minipage}{0.49\textwidth}
        \centering
        \caption*{(a) By Attack Method}
        \resizebox{\linewidth}{!}{
        \begin{tabular}{@{}lccccc@{}}
            \toprule
            \textbf{Method} & \textbf{N} & \textbf{Mean(\%)} & \textbf{Std(\%)} & \textbf{Min(\%)} & \textbf{Max(\%)} \\
            \midrule
            \textbf{Auto} & 7 & \textbf{64.70} & 19.87 & 36.52 & 91.09 \\
            C\&W & 14 & 53.89 & 18.94 & 27.76 & 85.16 \\
            DeepFool & 21 & 48.80 & 16.79 & 27.90 & 79.10 \\
            PGD & 21 & 44.63 & 15.85 & 23.20 & 71.76 \\
            BIM & 21 & 40.16 & 14.82 & 21.84 & 66.64 \\
            \textbf{FGSM} & 21 & \textbf{33.06} & 11.79 & 18.36 & 57.35 \\
            \bottomrule
        \end{tabular}
        }
    \end{minipage}
    \hfill
    \begin{minipage}{0.49\textwidth}
        \centering
        \caption*{(b) By Implementation Library}
        \resizebox{\linewidth}{!}{
        \begin{tabular}{@{}lccccc@{}}
            \toprule
            \textbf{Library} & \textbf{N} & \textbf{Mean(\%)} & \textbf{Std(\%)} & \textbf{Min(\%)} & \textbf{Max(\%)} \\
            \midrule
            ART & 42 & 48.52 & 19.12 & 18.36 & 91.09 \\
            Foolbox & 35 & 43.80 & 17.01 & 20.60 & 80.57 \\
            Custom & 28 & 40.58 & 15.57 & 19.97 & 75.53 \\
            \bottomrule
        \end{tabular}
        }
    \end{minipage}
    \vspace{0.5em}
    \begin{flushleft}
        \footnotesize \textit{Note: AutoAttack achieves the highest mean success rate (64.70\%), while FGSM shows the lowest but most consistent performance (lowest std: 11.79\%). ART library implementations demonstrate superior attack effectiveness.}
    \end{flushleft}
\end{table}

We first applied a Shapiro--Wilk normality test to compare the distribution of success rates across the six attack methods. We found that the results for FGSM, PGD, C\&W, DeepFool and AutoAttack were normally distributed, while only BIM deviated from normality. Even though BIM slightly violated normality, the sample sizes ($n \geq 21$) make ANOVA sufficiently robust to draw reliable conclusions.

In the homogeneity of variance test (Levene's test), we observed $W=1.1437$ and a $p$-value of $0.3425$. This indicates there were no significant differences among group variances (i.e., the assumption of homogeneity of variance was satisfied), making it appropriate to perform the ANOVA test.

\subsubsection{ANOVA and Variance Analysis}

Table~\ref{tab:anova_results} presents the comprehensive ANOVA analysis results:

\begin{table}[htbp]
    \centering
    \caption{ANOVA Analysis and Statistical Tests Results}
    \label{tab:anova_results}
    \resizebox{0.8\textwidth}{!}{
    \begin{tabular}{@{}lcccc@{}}
        \toprule
        \textbf{Statistical Test} & \textbf{Test Statistic} & \textbf{p-value} & \textbf{Effect Size} & \textbf{Interpretation} \\
        \midrule
        Levene's Test & W = 1.1437 & 0.3425 & --- & Homogeneity satisfied \\
        \textbf{One-way ANOVA} & \textbf{F(5,99) = 6.0593} & \textbf{0.0001***} & $\eta ^2 $ = 0.2343& \textbf{Large effect} \\
        \bottomrule
    \end{tabular}
    }
    \vspace{0.5em}
    \begin{flushleft}
        \footnotesize \textit{Note: ***p < 0.001 indicates highly significant differences. $\eta ^2 $ = 0.2343 represents a large effect size (Cohen's criterion: $\eta ^2\ \geq$  0.14). Attack method selection significantly impacts success rates.}
    \end{flushleft}
\end{table}

For the ANOVA test with $\alpha=0.05$, our results revealed a significant difference 
within the attack methods. We observed $F(5, 99)=6.0593$ with $p=0.0001$. The effect size 
($\eta^2 = 0.2343$) is also classified as a large effect according to Cohen ($\eta^2 \geq 
0.14$). This indicates that the selection of attack algorithm has a significant impact on 
the variance of attack success rates. 

\subsubsection{Multiple Comparison Analysis}

We also performed a multiple comparison using Tukey HSD, the results with significance are 
presented in Table~\ref{tab:tukey_hsd}, the full results are attached in Appendix~\ref{appendix:statistical_details}, along 
with the results of pairwise t-tests with Bonferroni correction for verification. 

\begin{table}[htbp]
    \centering
    \caption{Post-hoc Pairwise Comparison Using Tukey HSD (FWER = 0.05)}
    \label{tab:tukey_hsd}
    \resizebox{0.7\textwidth}{!}{
    \begin{tabular}{@{}llcccc@{}}
        \toprule
        \textbf{Group 1} & \textbf{Group 2} & \textbf{Mean Diff(\%)} & \textbf{p-value} & \textbf{Significant} & \textbf{Level}\\
        \midrule
        Auto     & FGSM      & 31.64 & 0.0002** & Yes & High\\
        Auto     & BIM       & 24.54 & 0.0076** & Yes & Moderate\\
        C\&W     & FGSM      & 20.82 & 0.0032** & Yes & Moderate\\
        DeepFool & FGSM      & 15.74 & 0.0210*  & Yes & Low\\
        \bottomrule
    \end{tabular}
}
    \vspace{0.5em}
    \begin{flushleft}
        \footnotesize
        \textit{Note: **p < 0.01, *p < 0.05. }
    \end{flushleft}
\end{table}

Effect size analysis results are presented in Table~\ref{tab:effect_size}. We observed that the largest effect was between FGSM and AutoAttack with Cohen’s $d = 2.249$, which is classified as a large effect. The comparisons of BIM with AutoAttack (Cohen's $d = 1.522$) and FGSM with C\&W (Cohen's $d = 1.386$) also show large effects. This indicates that these attack methods have significant differences in their attack success rates.

\begin{table}[htbp]
\centering
\caption{Effect Size (Cohen's \textit{d}) for Pairwise Comparisons}
\label{tab:effect_size}
\footnotesize

\begin{minipage}[t]{0.48\textwidth}
\centering
\begin{tabular}{@{}llcc@{}}
\toprule
\textbf{Group 1} & \textbf{Group 2} & \textbf{Cohen's \textit{d}} & \textbf{Effect Size} \\
\midrule
FGSM     & Auto      & 2.249 & Large \\
BIM      & Auto      & 1.522 & Large \\
FGSM     & C\&W      & 1.386 & Large \\
PGD      & Auto      & 1.190 & Large \\
FGSM     & DeepFool  & 1.085 & Large \\
DeepFool & Auto      & 0.906 & Large \\
FGSM     & PGD       & 0.829 & Large \\
C\&W     & BIM       & 0.828 & Large \\
\bottomrule
\end{tabular}
\end{minipage}
\hfill
\begin{minipage}[t]{0.48\textwidth}
\centering
\begin{tabular}{@{}llcc@{}}
\toprule
\textbf{Group 1} & \textbf{Group 2} & \textbf{Cohen's \textit{d}} & \textbf{Effect Size} \\
\midrule
C\&W     & Auto      & 0.562 & Medium \\
BIM      & DeepFool  & 0.546 & Medium \\
PGD      & C\&W      & 0.540 & Medium \\
FGSM     & BIM       & 0.530 & Medium \\
PGD      & BIM       & 0.292 & Small \\
C\&W     & DeepFool  & 0.288 & Small \\
PGD      & DeepFool  & 0.255 & Small \\
\bottomrule
\end{tabular}
\end{minipage}

\vspace{0.5em}
\begin{flushleft}
\footnotesize
\textit{Note: Effect size thresholds based on Cohen's convention — Small: $d \approx 0.2$, Medium: $d \approx 0.5$, Large: $d \geq 0.8$.}
\end{flushleft}
\end{table}

\subsubsection{Key Statistical Findings and Summary}

From our extensive numerical studies, we find the following important findings:

\begin{enumerate}
    \item \textbf{Attack Method Ranking}: AutoAttack achieved the highest attack success rate(64.70\%), with C\&W (53.89\%), DeepFool (48.80\%), PGD (44.63\%), BIM (40.16\%), FGSM (33.06\%) following behind.
    
    \item \textbf{Highly significant distinctions}: ANOVA test indicates significant differences between the methods ($F(5,99) = 6.06$, $p = 0.0001$) along with a large effect size ($\eta^2$ = 0.234).
    
    \item \textbf{FGSM Underperformance}: FGSM performance is significantly worse than AutoAttack ($p = 0.0002$), C\&W ($p = 0.0032$), and DeepFool ($p = 0.0210$) with very large to large effect sizes.
    
    \item \textbf{Impact on Implementation Libraries}: The ART library implementations result in the highest mean attack success rate (48.52\%), followed by Foolbox (43.80\%) and Custom implementations (40.58\%).
    
    \item \textbf{Practical significance}: The largest effect size observed between FGSM and AutoAttack (Cohen's $d=2.249$) corresponds to a substantial practical difference, underscoring the real-world importance of choosing robust attack methods. Comparisons of BIM with AutoAttack ($d=1.522$) and FGSM with C\&W ($d=1.386$) also reveal large practical differences.
\end{enumerate}

\begin{figure}[htbp]
    \centering
    \adjustbox{width=\textwidth,center}{%
        \includegraphics{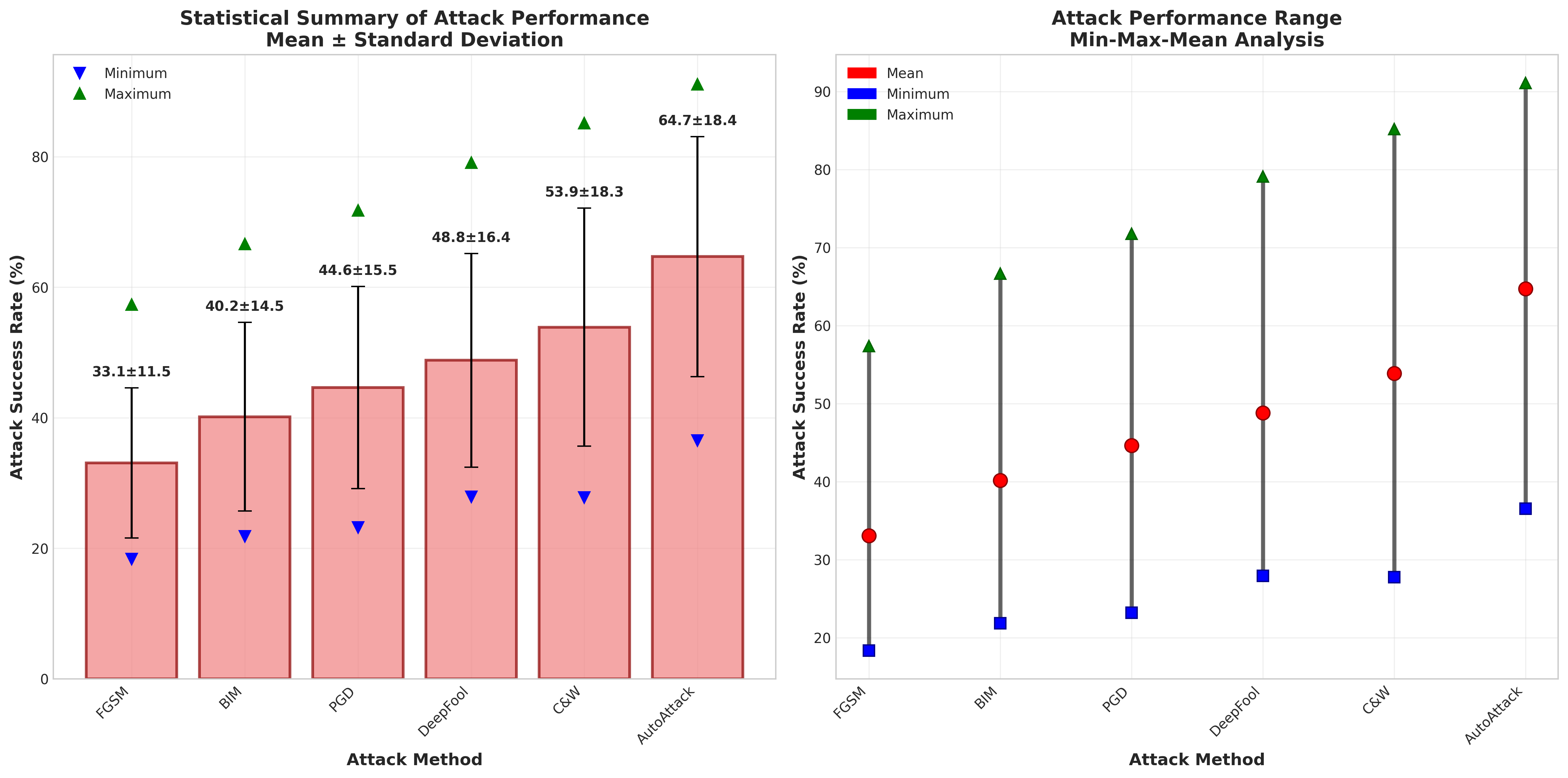}
    }
    \caption{Visualization of statistical analysis}
    \label{fig:statistical_summary}
\end{figure}

\subsection{Computational Efficiency Analysis}

The computational efficiency differences among various attack methods are significant, we present the computational time comparison in table~\ref{tab:computational_efficiency}:

\begin{table}[htbp]
    \centering
    \caption{Computational Efficiency Comparison}
    \label{tab:computational_efficiency}
    \resizebox{0.85\textwidth}{!}{
    \begin{tabular}{@{}lcccc@{}}
        \toprule
        \textbf{Method} & \textbf{Min Time (s)} & \textbf{Max Time (s)} & \textbf{Average Time (s)} & \textbf{Standard Deviation (s)} \\
        \midrule
        FGSM            & 0.050                 & 0.073                 & 0.056                     & 0.010                           \\
        PGD             & 0.754                 & 1.093                 & 0.900                     & 0.132                           \\
        C\&W            & 14.530                & 19.954                & 18.572                    & 2.040                           \\
        BIM             & 0.293                 & 0.390                 & 0.335                     & 0.052                           \\
        DeepFool        & 2.444                 & 3.146                 & 2.835                     & 0.407                           \\
        Auto            & 43.812                & 57.892                & 48.444                    & 4.550                           \\
        \bottomrule
    \end{tabular}
    }
\end{table}

\section{Discussion}

\subsection{Results Interpretation}

We evaluated a wide range of adversarial attacks on reinforcement learning-based medical questionnaire systems. Our results revealed several key vulnerabilities of such systems, we discuss them in the following sections.

\subsubsection{Attack Effectiveness Across Different Threat Models}

Our results demonstrate that the evaluated attack methods were all relatively effective, with ASR ranging from 33.1\% (FGSM) to 64.7\% (AutoAttack). Such high attack success rates could be especially concerning in clinical settings due to it's need for high recall rate and accuracy.  

Among the evaluated attack methods, AutoAttack achieved the maximum average ASR. This again demonstrated the effectiveness of ensemble attack methods, which aligned with the results from Croce et al.~\cite{croce2020reliable}, that such ensemble methods can provide a more comprehensive and effective adversarial robustness evaluation, compared to single attack methods.

Although AutoAttack did achieve the highest ASR, it also requires the most expensive computational resources compared to other attack methods. This demonstrates a tradeoff between the attack success rate and the attack efficiency. We classify the attack methods into three categories according to this tradeoff.
\begin{itemize}
    \item \textbf{High Success Rate Methods: }
    \begin{itemize}
        \item AutoAttack (Average ASR: 64.70\%, Computation Time: 47.094s)
        \item C\&W (Average ASR: 53.89\%, Computation Time: 18.194s)
    \end{itemize}
    \item \textbf{Balanced Methods: }
    \begin{itemize}
        \item DeepFool (Average ASR: 48.80\%, Computation Time: 2.778s)
        \item  PGD (Average ASR: 44.63\%, Computation Time: 0.880s)
    \end{itemize}
    \item \textbf{High Efficiency Methods: }
    \begin{itemize}
        \item FGSM (Average ASR: 33.06\%, Computation Time: 0.055s)
        \item  BIM (Average ASR: 40.16\%, Computation Time: 0.328s)
    \end{itemize}
\end{itemize}

\begin{figure}[htbp]
    \centering
    \adjustbox{width=0.8\textwidth,center}{%
        \includegraphics{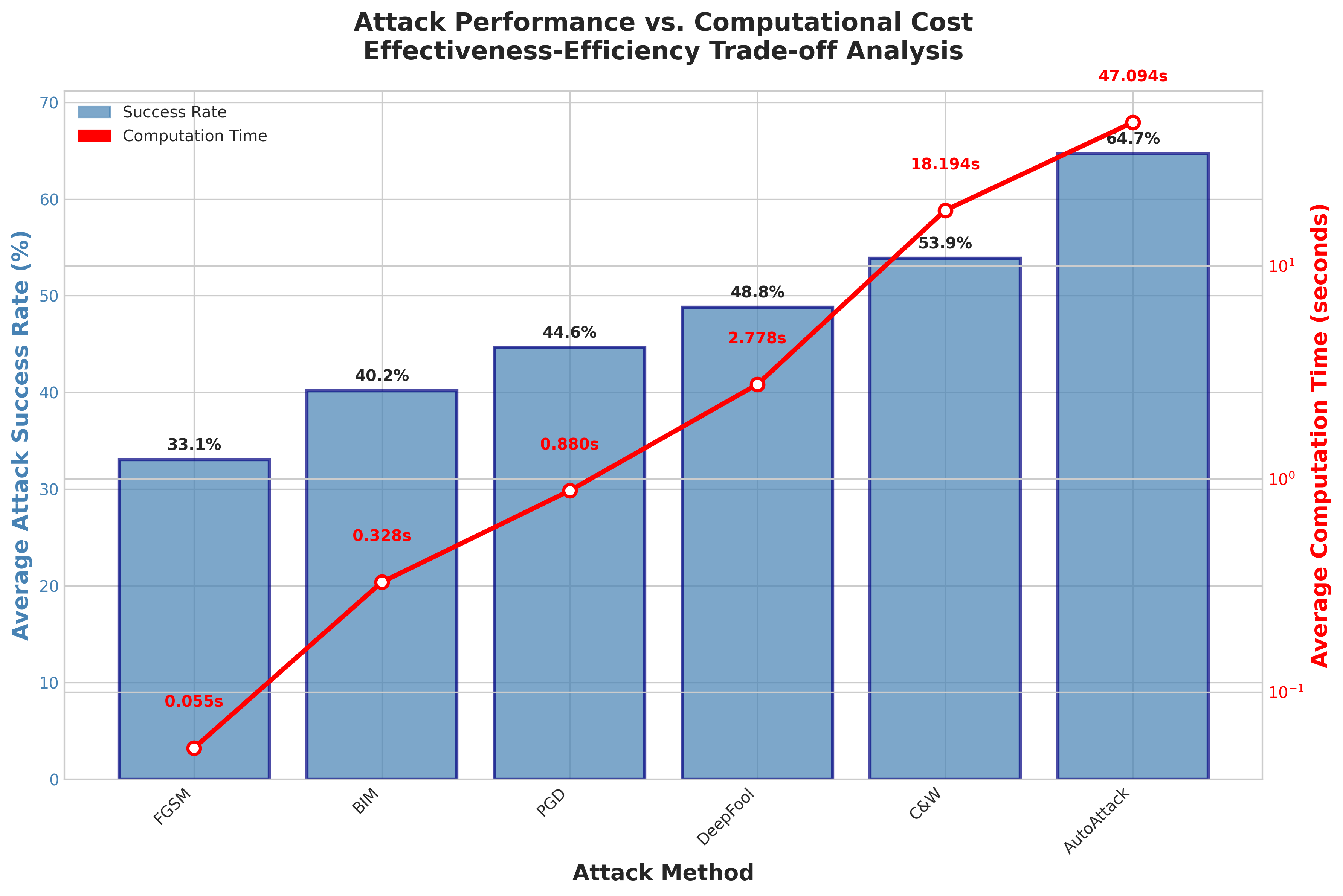}%
    }
    \caption{Computational time comparison across different attack methods. The results reveal a clear trade-off between computational efficiency and attack effectiveness.}
    \label{fig:computation_time}
\end{figure}

\subsubsection{Medical Constraint Framework Implications}

As stated before, we developed a medical constraint framework to generate clinically plausible adversarial attacks. It's high success rate (97.6\%) indicates that such examples could be generated with high efficiency. We suspect that this could make imperceptible adversarial attacks in clinical settings even more practical.

Another key limitation in generating medical adversarial attack examples proposed by Finlayson et al.~\cite{finlayson2019adversarial} is that previous attack methods frequently generated adversarial examples that did not fully represent patient data, and violated basic medical constraints. We resolved this problem by incorporating domain-specific knowledge into adversarial examples through solving a Constraint Satisfaction Problem (Algorithm~\ref{alg:constraint_satisfaction}).



\subsection{Comparison with Existing Literature}

Our results of the vulnerabilities of RL-based questionnaire systems align with the findings of Ma et al\cite{ma2020understanding} that AI image classification systems deployed in medical diagnosis settings are more vulnerable to adversarial attacks than such systems on the deployed for natural image classification. However, we further explored the generalizability of  such vulnerability to RL-based questionnaire systems, as they represent the application of sequential decision-making systems in medical settings. Such systems have a more complex attack surface, since the diagnosis policy are dynamic and relies on temporal information.

We also achieved a significantly higher success rate (Min: 33.1\%, Max: 64.7\%) compared with the results on medical image classification systems (Min: 15\%, Max: 25\%) reported by Finlayson et al.\cite{finlayson2019adversarial}.  This suggest that models trained to process tabular medical data is more vulnerable to adversarial attacks than that of medical images. We suspect this difference is due to the difference in data types. Since questionnaire systems are discrete while image pixels are continuous, the latter provides more attack vectors for manipulation. 

Although such vulnerability tends to be similar across tasks, it has domain-specific consequences, compared to prior work on adversarial attacks in RL~\cite{huang2017adversarial}. The strategic timing of attacks proposed by Lin et al.~\cite{lin2017tactics} is especially applicable to medical questionnaire systems, where the first few question selection mistakes have further impact throughout the entire diagnostic episode.

\subsection{Implications for Medical AI Safety}

\subsubsection{Clinical Deployment Risks}

The high adversarial attack success rates on such RL-based medical questionnaire systems could cause serious issues for their deployment in clinical settings. Also, since the generated adversarial examples remain imperceptible to clinicians, attackers can easily generate adversarial examples that are have a high success rate but are still hard to detect. This could pose another threat in practical deployment.

We suppose that successful adversarial attacks could potentially lead to:

\begin{itemize}
    \item \textbf{Delayed or incorrect diagnoses}: False negative diagnoses could delay certain medical treatments.
    \item \textbf{Unnecessary medical interventions}: False positives diagnoses may cause unnecessary treatments and associated costs.
    \item \textbf{Loss of clinician trust}: Repeated diagnostic errors may reduce clinician confidence in AI-assisted diagnosis.
    \item \textbf{System-wide vulnerabilities}: Successful attacks on one system may transfer to similar deployed systems.
\end{itemize}

\subsubsection{Regulatory and Ethical Considerations}
The ability to easily generate clinically plausible attacks suggests that current testing protocols are insufficient for detecting such attacks. Therefore, there is a critical need to develop a more advanced testing protocol, and also a need for enhanced regulations on medical AI systems. 

We thus call for the European Union (EU),  U.S. Food and Drugs Administration (FDA), and other institutions to include explicit requirements for the adversarial robustness of such systems in their respective laws or guidelines.

Such vulnerabilities of medical AI systems could also raise ethical problems. Patients should have the right to understand the limitations and potential risks of the vulnerabilities of AI-assisted diagnosis, and such systems should only be employed in clinical settings with the patient's consent.

\subsection{Limitations and Future Work}

Our study has several key limitations, we elaborate on these limitations from two perspectives: dataset and generalization limitations, and attack sophistication and practical considerations.

\subsubsection{Dataset and Generalization Limitations}

\begin{itemize}
    \item \textbf{Population Health Survey vs. Clinical Data}: The NHIS dataset used in this study is a population health survey instead of real-life clinical data. Therefore, it may not be able to fully represent the complexity of clinical settings.
    \item \textbf{Simplified Feature Space}: Due to limited computational resources, we only used the top-50 important features in our experiments. Such small number of features may fully represent the complexity of actual medical questionnaire systems.
    \item \textbf{Single Task Focus}: We only performed our evaluation on a single task, to predict the mortality rate of patients over a four-year period. However, real-life questionnaire systems usually handle multiple diagnostic tasks simultaneously. Our findings may not be able to generalize to such multi-task diagnosis systems.
\end{itemize}

Therefore, future research should further validate the our findings on multi-task diagnosis systems using appropriate clinical datasets. 

\subsubsection{Attack Sophistication and Practical Considerations}

\begin{itemize}
    \item \textbf{White-box Assumption}: We employed several white-box attacks in our evaluation framework. Such methods assume complete knowledge of model, including model gradients, parameters, and architecture. This data would be hardly be publicly available in real-life settings, making such methods particularly challenging to deploy. 
    \item \textbf{Perturbation Rates}: We also explored perturbation rates up to $\epsilon=2$, which is unlikely to be unnoticed in clinical settings. Such rates were only used to test the model's behaviors under worst-case scenarios. Thus, these perturbation levels may be impractical in real-life where the imperceptibility and subtlety of attacks are more important.
    \item \textbf{Detection Avoidance}: Although we deployed a medical constraint framework to ensure the clinical plausibility of generated adversarial examples, we did not take other detection mechanisms that might be deployed in real-life clinical systems into account (e.g. input sanitization or confidence-thresholds). 
\end{itemize}

\section{Conclusion}

Our work reviewed a wide range of adversarial attacks on Reinforcement learning-based adaptive medical questionnaire systems. We implemented and evaluated 7 distinct major attack strategies. Our results show that these systems are vulnerable to carefully crafted input perturbations generated by our medical constraint network. 

Our key contributions include: 
\begin{itemize}
    \item To the best of our knowledge, our work is the first comprehensive evaluation of adversarial attacks on reinforcement learning-based medical questionnaire systems. We demonstrated that such systems could be manipulated with different attacking methods, with attack success rates ranging from 33.1\% to 64.7\%.
    \item We developed a method-agnostic medical constraint network to generate adversarial examples that are clinically plausible. This framework reached a 97.6\% success rate in generating such examples. 
    \item We replicated the AdaptiveFS~\cite{shaham_learning_2020} model, and evaluated the previously mentioned methods using the NHIS datasets.
    \item We analyzed the significance of different methods when attacking RL-based medical questionnaire systems. We also propose that the difference in implementation can also have impact on the attack success rates.
\end{itemize}

Our findings suggest that adversarial robustness should be considered as an important requirement for medical AI systems' deployment. The high success rates of the evaluated attack methods suggests that the AdaptiveFS framework~\cite{shaham_learning_2020} used in this study could have critical underlying vulnerabilities. To the best of our knowledge, no fix has been proposed to prevent the generalization of such attacks across different RL-based medical AI systems. Therefore, our work aims to resolve this problem by providing a comprehensive evaluation for the model's robustness and reliability. 

We also found that domain-specific constraint frameworks (e.g. our medical constraint framework) could be used to generate plausible adversarial examples, and thus help the attacks remain imperceptible to humans. The strong generalizability of this method could be applied in different scenarios, which again demonstrates the urgent need for a robust and precise attack detection mechanism.

The findings from our work can provides several suggestions for healthcare providers and AI developers. We call for the healthcare providers to perform a comprehensive adversarial robustness evaluation of the system, and grant patients' consent before deploying such systems in clinical settings. Our work pointed out several vulnerabilities of current systems as mentioned in previous sections. These vulnerabilities could be resolved by the AI developers, thus providing a more robust system. Our current solution includes applying adversarial robustness testing on the system level, continuous monitoring in deployment, and deploying domain-specific validation frameworks on the input level.

Although our current work has several limitations, it still serves as a foundation for future research on the adversarial robustness of RL-based medical questionnaire systems. The evaluation framework proposed in this work could be extended to broader applications, including multi-module AI diagnosis systems with questionnaires, general RL-based questionnaire systems, etc. Our work can contribute to the development of a more generalizable evaluation framework.


\newpage

\appendix

\section{Medical Constraint Framework Details}
\label{appendix:constraint_details}

\subsection{Detailed Constraint Violation Examples and Corrections}

Our medical constraint framework encountered various violation patterns during the generation of adversarial examples. Below are representative examples showing how violations were detected and corrected:

\begin{itemize}
    \item \textbf{Age-BMI Violation}: 
    \begin{itemize}
        \item \textit{Original}: Age=25, BMI=35 → \textit{Perturbed}: Age=25, BMI=18
        \item \textit{Issue}: Sudden weight loss from obese to underweight is medically implausible
        \item \textit{Correction}: BMI clamped to 22.5 (minimum healthy BMI for young adults)
    \end{itemize}
    
    \item \textbf{Pregnancy-Gender Conflict}: 
    \begin{itemize}
        \item \textit{Original}: Male, Not Pregnant → \textit{Perturbed}: Male, Pregnant
        \item \textit{Issue}: Biological impossibility
        \item \textit{Correction}: Pregnancy status reset to "Not Pregnant" while maintaining other perturbations
    \end{itemize}
    
    \item \textbf{Diabetic-Glucose Inconsistency}: 
    \begin{itemize}
        \item \textit{Original}: Diabetic, Glucose=180mg/dL → \textit{Perturbed}: Diabetic, Glucose=80mg/dL
        \item \textit{Issue}: Diagnosed diabetic with normal glucose levels without medication
        \item \textit{Correction}: Glucose adjusted to 140mg/dL (lower bound for diabetic patients)
    \end{itemize}
    
    \item \textbf{Smoking-Lung Disease Correlation}: 
    \begin{itemize}
        \item \textit{Original}: Non-smoker, No COPD → \textit{Perturbed}: Non-smoker, Severe COPD
        \item \textit{Issue}: Severe COPD in non-smoker without environmental exposure
        \item \textit{Correction}: Either smoking status changed to "Former smoker" or COPD severity reduced to "Mild"
    \end{itemize} 
\end{itemize}

\subsection{Constraint Satisfaction Algorithm Performance}

The constraint satisfaction algorithm showed the following characteristics:

\begin{itemize}
    \item \textbf{Automatic Resolution}: 94.2\% of conflicts resolved automatically through constraint projection
    \item \textbf{Iterative Refinement}: 4.1\% required iterative refinement (average 2.3 iterations)
    \item \textbf{Irreconcilable Violations}: 1.7\% rejected due to irreconcilable violations (e.g., 90-year-old with fertility-related perturbations)
    \item \textbf{Convergence Time}: Average 0.23 seconds per constraint satisfaction operation
    \item \textbf{Rule Coverage}: 247 total rules across 5 categories (physiological bounds, correlations, conditional constraints, temporal consistency, demographic validity)
\end{itemize}

\section{Epsilon Medical Validation Details}
\label{appendix:epsilon_validation}

\subsection{Concrete Examples of $\epsilon$=2.0 Medical Plausibility}

For $\epsilon$=2.0 perturbations in the normalized [-1,1] space, we provide concrete examples demonstrating medical plausibility:

\begin{itemize}
    \item \textbf{Age perturbation}: $\epsilon$=0.3 in normalized space
    \begin{itemize}
        \item Range: [18, 85] years → Normalized: [-1, 1]
        \item Perturbation: 0.3 × (85-18)/2 = 10.05 years
        \item Example: 45 → 55 years (realistic aging or measurement uncertainty)
    \end{itemize}
    
    \item \textbf{BMI adjustment}: $\epsilon$=0.5 in normalized space
    \begin{itemize}
        \item Range: [15, 45] kg/m² → Normalized: [-1, 1]
        \item Perturbation: 0.5 × (45-15)/2 = 7.5 kg/m²
        \item Example: 26 → 33.5 kg/m² (weight gain or measurement variation)
    \end{itemize}
    
    \item \textbf{Blood pressure}: $\epsilon$=0.4 in normalized space
    \begin{itemize}
        \item Range: [80, 200] mmHg → Normalized: [-1, 1]
        \item Perturbation: 0.4 × (200-80)/2 = 24 mmHg
        \item Example: 120 → 144 mmHg (stress-induced elevation)
    \end{itemize}
    
    \item \textbf{Cumulative multi-dimensional effect}: 
    \begin{itemize}
        \item $L \infty=2.0$ allows simultaneous moderate changes across multiple features
        \item Example: Age +5 years, BMI +2 units, BP +15 mmHg, Glucose +20 mg/dL
        \item Represents gradual health deterioration or lifestyle changes over time
    \end{itemize}
\end{itemize}

\section{Attack Parameter Settings}
\label{appendix:attack_params}

\begin{table}[htbp]
    \centering
    \caption{Complete Attack Parameter Settings for Experimental Evaluation}
    \resizebox{0.8\textwidth}{!}{
    \begin{tabular}{@{}lcccc@{}}
        \toprule
        \textbf{Attack Method} & \textbf{$\epsilon$ Values} & \textbf{Norm} & \textbf{Iterations} & \textbf{Other Parameters} \\
        \midrule
        FGSM                   & [0.1, 0.3, 0.5, 0.8, 1.0, 1.5, 2.0]            & $L_\infty$ & 1                   & ---                       \\
        PGD ($L_\infty$)       & [0.1, 0.3, 0.5, 0.8, 1.0, 1.5, 2.0]            & $L_\infty$ & 40                  & $\alpha = \epsilon/40$    \\
        PGD ($L_2$)            & [0.1, 0.3, 0.5, 0.8, 1.0, 1.5, 2.0]            & $L_2$      & 40                  & $\alpha = \epsilon/40$    \\
        C\&W                   & [0.1, 0.3, 0.5, 0.8, 1.0, 1.5, 2.0]            & $L_2$      & 100                 & $\kappa = 0$, $c = 1e-4$ \\
        BIM                    & [0.1, 0.3, 0.5, 0.8, 1.0, 1.5, 2.0]            & $L_\infty$ & 40                  & $\alpha = \epsilon/40$    \\
        DeepFool               & [0.1, 0.3, 0.5, 0.8, 1.0, 1.5, 2.0]            & $L_2$      & 100                 & overshoot = 0.02          \\
        AutoAttack             & [0.1, 0.3, 0.5, 0.8, 1.0, 1.5, 2.0]            & Mixed      & Variable            & ensemble of FGSM+PGD+C\&W \\
        \bottomrule
    \end{tabular}
    }
\end{table}

\section{Feature Details}
\label{appendix:feature_details}

\subsection{Complete Feature Set Used in Experiments}

Following the AdaptiveFS framework~\cite{shaham_learning_2020}, our experiments utilized the 50 most important features selected by XGBoost importance ranking from the full NHIS dataset. This feature selection approach is consistent with the original AdaptiveFS methodology and ensures optimal performance for the reinforcement learning-based questionnaire system. Table~\ref{tab:nhis_features} presents the complete set of 50 features used in all adversarial attack experiments.

\begin{table}[htbp]
\centering
\caption{Complete Set of 50 NHIS Features Used in Adversarial Attack Experiments}
\label{tab:nhis_features}
\footnotesize

\begin{minipage}[t]{0.48\textwidth}
\centering
\begin{tabular}{@{}p{2.5cm}p{4.5cm}@{}}
\toprule
\textbf{Feature Name} & \textbf{Description} \\
\midrule
medicare1 & Medicare coverage recode \\
la1ar1 & Any limitation - all persons, all conditions \\
flwalk0 & How difficult to walk 1/4 mile without special equipment \\
age-p & Age \\
flclimb0 & How difficult to climb 10 steps without special equipment \\
doinglwp5 & What was patient doing last week \\
la1ar2 & Any limitation - all persons, all conditions \\
flcarry0 & How difficult to lift/carry 10 lbs without special equipment \\
wrklyr12 & Work for pay last year \\
pregnow999 & Currently pregnant \\
smkev1 & Ever smoked 100 cigarettes \\
lupprt1 & Lost all upper and lower natural teeth \\
phstat5 & Reported health status \\
speceq2 & Have health problem that requires special equipment \\
flshop0 & How difficult to go out to events without special equipment \\
flwalk4 & How difficult to walk 1/4 mile without special equipment \\
fliadlyn2 & Any family member need help with an IADL \\
smkev2 & Ever smoked 100 cigarettes \\
educ15 & Highest level of school completed \\
phstat4 & Reported health status \\
eligpwic & Anyone age-eligible for the WIC program \\
canev1 & Ever told by a doctor you had cancer \\
adnlong21 & Time since last saw a dentist \\
vigfreqw & Freq vigorous activity (times per week) \\
sex & Sex \\
\bottomrule
\end{tabular}
\end{minipage}
\hfill
\begin{minipage}[t]{0.48\textwidth}
\centering
\begin{tabular}{@{}p{2.5cm}p{4.5cm}@{}}
\toprule
\textbf{Feature Name} & \textbf{Description} \\
\midrule
livyr2 & Told you had liver condition, past 12 months \\
private2 & Private health insurance recode \\
ahchyr1 & Received home care from health professional, past 12 months \\
ahcsyr71 & Seen/talked to mental health professional, past 12 months \\
dibev1 & Ever been told that you have diabetes \\
ephev1 & Ever been told you had emphysema \\
miev1 & Ever been told you had a heart attack \\
kidwkyr2 & Told you had weak/failing kidneys, past 12 months \\
phstat1 & Reported health status \\
flsocl0 & How difficult to participate in social activities without special equipment \\
phstat2 & Reported health status \\
ahchyr2 & Received home care from health professional, past 12 months \\
hiscodi32 & Race/ethnicity recode \\
livyr1 & Told you had liver condition, past 12 months \\
bmi & Body Mass Index (BMI) \\
amigr2 & Had severe headache/migraine, past 3 months \\
rat-cat24 & Ratio of family income to the poverty threshold \\
jntsymp1 & Symptoms of joint pain/aching/stiffness past 30 days \\
houseown2 & Home tenure status \\
doinglwp1 & What was patient doing last week \\
beddayr & Number of bed days, past 12 months \\
ahernoy2 & Times in ER/ED, past 12 months \\
proxysa2 & Sample adult status \\
\bottomrule
\end{tabular}
\end{minipage}

\vspace{0.5em}
\begin{flushleft}
\footnotesize \textit{Note: These 50 features were selected based on XGBoost importance ranking from the full NHIS dataset containing 1,182 total features. The selection methodology follows the AdaptiveFS framework~\cite{shaham_learning_2020} to ensure optimal performance for reinforcement learning-based medical questionnaire systems.}
\end{flushleft}
\end{table}

This feature set represents the core variables used throughout our adversarial attack evaluation, encompassing demographic information, health status indicators, functional limitations, medical conditions, lifestyle factors, and healthcare utilization patterns. All adversarial perturbations and medical constraint validations were applied specifically to these 50 features to ensure clinical relevance and experimental consistency with the original AdaptiveFS methodology.

\section{Training Configuration Details}
\label{appendix:training_details}

\subsection{Complete Training Configuration}

The AdaptiveFS framework training employed the following detailed configuration:

\begin{itemize}
    \item \textbf{Learning Rate Schedule}: 
    \begin{itemize}
        \item Initial rate: $\eta = 10^{-4}$
        \item Decay schedule: Step-wise reduction by factor 0.1 every 17,500 steps
        \item Minimum rate: $1 \times 10^{-6}$
    \end{itemize}
    
    \item \textbf{Validation Protocol}:
    \begin{itemize}
        \item Frequency: Every 1,000 episodes
        \item Early stopping: 50 validation trials without improvement
        \item Metric: Validation AUC (primary), accuracy (secondary)
        \item Validation set: 5\% of training data, max 20,000 samples
    \end{itemize}
    
    \item \textbf{Reward Function}:
    \begin{itemize}
        \item Diagnostic guess: $R = p(y_{true}|s)$ (model confidence for correct class)
        \item Intermediate steps: Small random reward $\sim \mathcal{N}(0, 0.01)$
        \item Episode termination: +1 for correct diagnosis, -1 for incorrect
    \end{itemize}
    
    \item \textbf{Experience Replay}:
    \begin{itemize}
        \item Buffer size: 1,000 transitions
        \item Sampling: Uniform random
        \item Update frequency: Every 4 steps
        \item Batch size: 32 transitions
    \end{itemize}
    
    \item \textbf{Training Schedule}:
    \begin{itemize}
        \item Alternating training: DQN and Guesser networks
        \item Switch frequency: Every 1,000 episodes
        \item Total episodes: Up to 50,000 with early stopping
    \end{itemize}
    
    \item \textbf{Network Architecture Details}:
    \begin{itemize}
        \item DQN: 128-dimensional hidden layers, ReLU activation
        \item Guesser: 256-dimensional hidden layers, PReLU activation
        \item Dropout: 0.1 during training
        \item Weight initialization: Xavier uniform
    \end{itemize}
    
    \item \textbf{Target Network Updates}:
    \begin{itemize}
        \item Update frequency: Every 10 episodes
        \item Update method: Hard copy ($\tau = 1.0$)
        \item Target freezing: 100 episodes for stability
    \end{itemize}
\end{itemize}

\section{Implementation Details}
\label{appendix:implementation_details}

\subsection{Computational Resources}

\begin{itemize}
    \item \textbf{Hardware Configuration}:
    \begin{itemize}
        \item GPU: NVIDIA A100 (40GB VRAM)
        \item RAM: 64GB
        \item Storage: 1TB NVMe SSD
    \end{itemize}
    
    \item \textbf{Software Environment}:
    \begin{itemize}
        \item OS: Ubuntu 20.04 LTS
        \item Python: 3.8.10
        \item PyTorch: 2.7.1
        \item CUDA: 12.6
        \item Additional libraries: NumPy 1.21.0, Pandas 1.3.0, Scikit-learn 1.0.2
    \end{itemize}
    
    \item \textbf{Attack Libraries}:
    \begin{itemize}
        \item Adversarial Robustness Toolbox (ART): v1.15.1
        \item Foolbox: v3.3.3
        \item Custom implementations for method-specific optimizations
    \end{itemize}
\end{itemize}

\subsection{Dataset Configuration Details}

\begin{itemize}
    \item \textbf{NHIS Dataset Specifications}:
    \begin{itemize}
        \item Total observations: 182,630 across 7 years (2005-2011)
        \item Total features: 1,182 (Case 200 configuration: 50 core features)
        \item Temporal split: 2005-2009 (122,019 samples) training, 2010-2011 (60,611 samples) testing
        \item Mortality rate: 4.5\% (8,131 deaths over 4-year follow-up)
    \end{itemize}
    
    \item \textbf{Preprocessing Pipeline}:
    \begin{itemize}
        \item Normalization: Min-max scaling to [-1, 1] range
        \item Missing value imputation: Median for continuous, mode for categorical
        \item Categorical encoding: One-hot encoding followed by normalization
        \item Feature selection: Correlation-based removal (threshold 0.95)
    \end{itemize}
    
    \item \textbf{Evaluation Configuration}:
    \begin{itemize}
        \item Attack evaluation samples: 1,000 correctly classified from test set across 42 experimental configurations
        \item Statistical power: 0.8 for effect size Cohen's $d \geq 0.18$
        \item Significance level: $\alpha = 0.05$
        \item Confidence intervals: 95\% (±0.031 for success rates)
        \item Cross-validation: 5-fold for hyperparameter tuning
    \end{itemize}
    
    \item \textbf{Episode Configuration}:
    \begin{itemize}
        \item Maximum episode length: 8 questions
        \item Average episode length: 4.2 questions (clean inputs)
        \item Question selection strategy: $\epsilon$-greedy with decay
        \item State representation: [features, question\_mask] $\in \mathbb{R}^{100}$
    \end{itemize}
\end{itemize}

\section{Statistical Analysis Details}
\label{appendix:statistical_details}

\begin{table}[htbp]
    \centering
    \caption{Complete Pairwise Comparisons (Tukey HSD)}
    \label{tab:tukey_complete}
    \resizebox{0.7\textwidth}{!}{
    \begin{tabular}{@{}llcccc@{}}
        \toprule
        \textbf{Group 1} & \textbf{Group 2} & \textbf{Mean Diff} & \textbf{95\% CI Lower} & \textbf{95\% CI Upper} & \textbf{p-value} \\
        \midrule
        AutoAttack & FGSM & 31.64 & 11.54 & 51.74 & 0.0002** \\
        AutoAttack & BIM & 24.54 & 4.44 & 44.65 & 0.0076** \\
        AutoAttack & PGD & 20.07 & 0.04 & 40.17 & 0.0507 \\
        AutoAttack & DeepFool & 15.90 & 4.20 & 36.01 & 0.2044 \\
        AutoAttack & C\&W & 10.82 & 10.51 & 32.14 & 0.6813 \\
        C\&W & FGSM & 20.82 & 4.93 & 36.72 & 0.0032** \\
        C\&W & BIM & 13.73 & 2.17 & 29.62 & 0.1312 \\
        C\&W & PGD & 9.25 & 6.64 & 25.14 & 0.5408 \\
        C\&W & DeepFool & 5.08 & 10.81 & 20.98 & 0.9379 \\
        DeepFool & FGSM & 15.74 & 1.52 & 29.95 & 0.0210* \\
        DeepFool & BIM & 8.64 & 5.57 & 22.86 & 0.4918 \\
        DeepFool & PGD & 4.16 & 10.05 & 18.38 & 0.9569 \\
        PGD & FGSM & 11.57 & 2.64 & 25.79 & 0.1783 \\
        PGD & BIM & 4.48 & 9.74 & 18.69 & 0.9417 \\
        BIM & FGSM & 7.10 & 7.12 & 21.31 & 0.6958 \\
        \bottomrule
    \end{tabular}
    }
\end{table}

\begin{table}[htbp]
    \centering
    \caption{Pairwise t-tests with Bonferroni Correction ($\alpha = 0.0033$)}
    \label{tab:bonferroni_ttest}
    \resizebox{0.7\textwidth}{!}{
    \begin{tabular}{@{}llccccc@{}}
        \toprule
        \textbf{Group 1} & \textbf{Group 2} & \textbf{Mean Diff} & \textbf{t} & \textbf{p-value} & \textbf{Cohen's \textit{d}} & \textbf{Sig} \\
        \midrule
        FGSM     & PGD       & -0.1157 & -2.685 & 0.0105 & -0.829 &  \\
        FGSM     & C\&W      & -0.2082 & -4.018 & 0.0003 & -1.386 & *** \\
        FGSM     & BIM       & -0.0710 & -1.717 & 0.0936 & -0.530 &  \\
        FGSM     & DeepFool  & -0.1574 & -3.516 & 0.0011 & -1.085 & *** \\
        FGSM     & Auto      & -0.3164 & -5.152 & 0.0000 & -2.249 & *** \\
        PGD      & C\&W      & -0.0925 & -1.564 & 0.1273 & -0.540 &  \\
        PGD      & BIM       &  0.0448 &  0.945 & 0.3501 &  0.292 &  \\
        PGD      & DeepFool  & -0.0416 & -0.826 & 0.4134 & -0.255 &  \\
        PGD      & Auto      & -0.2007 & -2.726 & 0.0113 & -1.190 &  \\
        C\&W     & BIM       &  0.1373 &  2.401 & 0.0221 &  0.828 &  \\
        C\&W     & DeepFool  &  0.0508 &  0.834 & 0.4103 &  0.288 &  \\
        C\&W     & Auto      & -0.1082 & -1.215 & 0.2394 & -0.562 &  \\
        BIM      & DeepFool  & -0.0864 & -1.768 & 0.0846 & -0.546 &  \\
        BIM      & Auto      & -0.2454 & -3.487 & 0.0018 & -1.522 & *** \\
        DeepFool & Auto      & -0.1590 & -2.076 & 0.0479 & -0.906 &  \\
        \bottomrule
    \end{tabular}
    }
    \vspace{0.5em}
    \begin{flushleft}
        \footnotesize
        \textit{Note: Significance threshold is Bonferroni-corrected $\alpha = 0.0033$. Values marked with \textbf{***} are significant after correction.}
    \end{flushleft}
\end{table}

\newpage
\bibliographystyle{ieeetr}
\bibliography{ref}

\end{document}